\pdfoutput=1
\documentclass[prl,10pt,twocolumn,reprint,showkeys,superscriptaddress,nofootbib,longbibliography]{revtex4-1}

\usepackage{amsmath}    
\usepackage{graphicx}   
\usepackage{verbatim}   
\usepackage{color}      
\usepackage{subfigure}  
\usepackage{hyperref}   
\usepackage{SIunits}     
\usepackage{braket}

\begin{document}

\title{Measurement and Control of Single Nitrogen-Vacancy Center Spins above 600\,K}
\author{D. M. Toyli}
 \thanks{These two authors contributed equally to this work.}
 \affiliation{Center for Spintronics and Quantum Computation,\\ University of
California, Santa Barbara, CA 93106-6105, USA}
\author{D. J. Christle}
 \thanks{These two authors contributed equally to this work.}
 \affiliation{Center for Spintronics and Quantum Computation,\\ University of
California, Santa Barbara, CA 93106-6105, USA}
\author{A. Alkauskas}
\affiliation{Center for Spintronics and Quantum Computation,\\ University of
California, Santa Barbara, CA 93106-6105, USA}
\affiliation{Materials Department,\\ University of
California, Santa Barbara, CA 93106-5050, USA}
\author{B.\ B.\ Buckley}
\affiliation{Center for Spintronics and Quantum Computation,\\ University of
California, Santa Barbara, CA 93106-6105, USA}
\author{C. G. Van de Walle}
\affiliation{Center for Spintronics and Quantum Computation,\\ University of
California, Santa Barbara, CA 93106-6105, USA}
\affiliation{Materials Department,\\ University of
California, Santa Barbara, CA 93106-5050, USA}
\author{D.\ D.\ Awschalom}
\email[Corresponding author.\\ Email address: ]{awsch@physics.ucsb.edu}
\affiliation{Center for Spintronics and Quantum Computation,\\ University of
California, Santa Barbara, CA 93106-6105, USA}
\date{\today}
\begin{abstract}We study the spin and orbital dynamics of single nitrogen-vacancy (NV) centers in diamond between room temperature and \unit{700}{\kelvin}. We find that the ability to optically address and coherently control single spins above room temperature is limited by nonradiative processes that quench the NV center's fluorescence-based spin readout between \unit{550}{\kelvin} and \unit{700}{\kelvin}. Combined with electronic structure calculations, our measurements indicate that the energy difference between the $^{3}E$ and $^{1}A_{1}$ electronic states is $\sim\!0.8\,\mathrm{eV}$.  We also demonstrate that the inhomogeneous spin lifetime ($T_2^{*}$) is temperature independent up to at least \unit{625}{\kelvin}, suggesting that single NV centers could be applied as nanoscale thermometers over a broad temperature range.
\end{abstract}
\maketitle
\SIunits[thinspace,thinqspace]
The negatively-charged nitrogen-vacancy (NV) center spin in diamond stands out among individually-addressable qubit systems because it can be initialized, coherently controlled, and read out at room temperature \cite{Jelezko_PRL_2004}. The defect's robust spin coherence \cite{Balasubramanian_NatureMaterials_2009} and optical addressability via spin-dependent orbital transitions \cite{Wrachtrup_Science_1997} have enabled applications ranging from quantum information processing \cite{Dutt_Science_2007,Neumann_Science_2010,Buckley_Science_2010,Robledo_Nature_2011} to nanoscale magnetic and electric field sensing \cite{Maze_Nature_2008,Balasubramanian_Nature_2008,Dolde_NatPhys_2011}.  While it has been shown that NV center spins can be optically polarized up to at least \unit{500}{\kelvin} \cite{Loubser_DR_1977,Redman_PRL_1991}, little is known about what processes limit the spin's optical addressability and coherence at higher temperatures. Understanding these processes is important to high-temperature field sensing applications and will aid the search for new defect-based spin qubits analogous to the NV center \cite{Weber_PNAS_2010,Koehl_Nature_2011} by identifying the aspects of its orbital structure responsible for its high temperature operation.

\begin{figure}
  \includegraphics[width=\columnwidth]{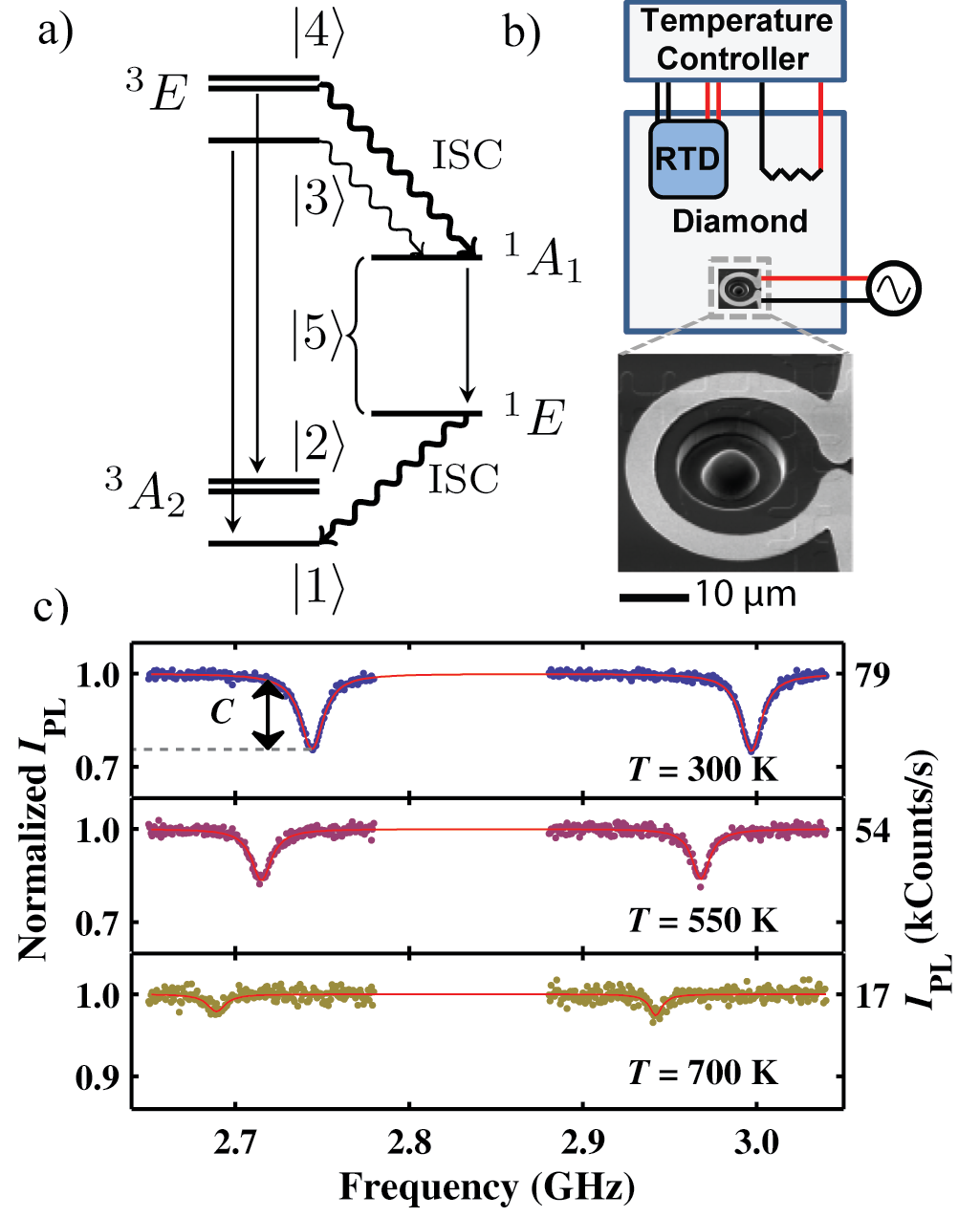}\\
  \caption{\label{fig1}(color online) (a) Orbital diagram showing the triplet ($^{3}A_2$ and $^{3}E$) and singlet ($^{1}A_{1}$ and $^{1}E$) electronic states. The straight lines represent optical transitions and the snaked lines represent nonradiative intersystem crossings (ISC). Bra-ket notation denotes the effective levels used to model the photoluminescence intensity ($I_\mathrm{PL}$) and electron spin resonance (ESR) contrast.  Level $\ket{2}$ ($\ket{4}$) is a grouping of the $^{3}A_2$ ($^{3}E$) $m_s = \pm 1$ sublevels and $\ket{5}$ is an effective singlet level. (b) Sample schematic showing the resistive heater and resistive temperature detector (RTD), the solid immersion lens (SIL), and the antenna. The scanning electron microscope image shows the SIL and antenna.  (c) $I_\mathrm{PL}$ of a single NV center versus applied microwave frequency. Lorentzian dips with normalized amplitude $C$ are observed at the spin resonance frequencies ($B \sim\!45\,\mathrm{G}$). The left (right) axis shows the normalized (absolute) $I_\mathrm{PL}$ to illustrate the temperature-dependent decrease in $C$ and $I_\mathrm{PL}$.  The red lines are two-Lorentzian fits.  The crystal field splitting ($D$) is the average of the resonance frequencies. The measurements were performed at high microwave power to saturate the spin transitions.}
\end{figure}

The NV center's optical spin polarization and optical spin readout result from a spin-selective intersystem crossing (ISC). Although optical transitions between the spin-triplet ground ($^{3}A_{2}$) and excited ($^{3}E$) states ($1.945\,\mathrm{eV}$ zero-phonon line [ZPL]) are typically spin conserving, the $^{3}E$ state can also relax to the $^{3}A_{2}$ state via an indirect pathway that involves a nonradiative, triplet to singlet ISC and subsequent transitions through at least one additional singlet [Fig. \ref{fig1}(a)].  The $^{3}E$ ISC is much stronger for the $m_s = \pm 1$ $^{3}E$ sublevels than for the $m_s = 0$ sublevel, which facilitates spin readout through the resulting spin-dependent photoluminescence (PL) and initializes the spin into the $m_s = 0$ $^{3}A_{2}$ sublevel with high probability ($P_{m_s = 0} \sim\!0.8$) through repeated optical excitation \cite{Manson_PRB_2006}.  Despite the singlet pathway's critical role in preparing and interrogating the spin, open questions remain regarding the number and energies of the singlets involved.  Recent experiments have established that it consists of at least two singlets of $^{1}A_{1}$ and $^{1}E$ symmetry separated by $1.19\,\mathrm{eV}$ \cite{Rogers_NJP_2008,Acosta_PRB_2010,Manson_arXiv_2010}, and have shown that the $^{1}E$ state lifetime is strongly temperature dependent below \unit{300}{\kelvin} \cite{Acosta_PRB_2010,Robledo_NJP_2011}. The details of the $^{3}E$ ISC remain uncertain, however, given computations that suggest the singlet pathway involves three singlets \cite{Gali_PRB_2010} and the lack of direct measurements of the $^{3}E$ to singlet transition energy.

In this work, we report measurements of single NV center spins between \unit{300}{\kelvin} and \unit{700}{\kelvin}. We show that thermally-activated nonradiative processes diminish the spin-selectivity of the $^{3}E$ ISC and quench the optical spin readout above \unit{550}{\kelvin}. Our measurements indicate the energy barrier for nonradiative relaxation from the $^{3}E$ state is $\sim\!0.5\,\mathrm{eV}$. We perform electronic structure calculations that suggest an orbital structure consistent with a two-singlet decay pathway, the measured energy barrier for nonradiative relaxation, and the temperature-dependence of the $^{1}E$ state lifetime. Furthermore, we demonstrate that the spin coherence remains robust even as nonradiative processes quench the optical spin readout. In addition to demonstrating the defect's potential for high-temperature magnetic field sensing applications, these results suggest single NV centers could act as nanoscale thermometers with sensitivities on the order of $100\,\mathrm{mK}/\sqrt{\mathrm{Hz}}$ between room temperature and \unit{600}{\kelvin}.  This high sensitivity and wide range of operating temperatures make NV centers an attractive candidate for a variety of thermosensing applications such as cellular thermometry \cite{Yang_ACSNano_2011} and diamond-based scanning thermal microscopy \cite{Majumdar_RevSciInst_1995}.

To study single NV centers at elevated temperatures, we constructed devices that combine on-chip heating and thermometry elements, solid immersion lenses (SILs) for enhanced photon collection efficiency \cite*{Hadden_APL_2010,Marseglia_APL_2011}, and on-chip microwave elements for spin control [Fig. \ref{fig1}(b)] \cite{Fuchs_Science_2009,SOM}. We studied NV centers in a single-crystal diamond sample ($[\mathrm{N}]_\mathrm{S}^{0} < 5\,\mathrm{ppb}$) using a home-built confocal microscope \cite{Epstein_NatPhys_2005}. We etched hemispherical SILs around single defects with a focused ion beam, patterned microwave antennas around the NV centers, and patterned a resistive heater on the sample surface. We adhered a commercial resistive temperature detector (RTD) to the diamond and connected the heater and RTD to a temperature controller to achieve stability within $\pm50\,\mathrm{mK}$ up to \unit{700}{\kelvin} in air.

Continuous-wave (CW) electron spin resonance (ESR) measurements revealed that the NV center's PL intensity ($I_\mathrm{PL}$) and relative $I_\mathrm{PL}$ difference between its spin states (ESR contrast) strongly decrease above \unit{550}{\kelvin}. The measurements were carried out by applying a swept-frequency AC magnetic field while monitoring $I_\mathrm{PL}$ under continuous \unit{532}{\nano\meter} laser excitation.  We observed Lorentzian dips in $I_\mathrm{PL}$ with normalized amplitude $C$ centered at the ground-state spin resonance frequencies \cite{Wrachtrup_Science_1997}.  The resonance frequencies are determined by the crystal field splitting ($D$) between the $m_s = \pm 1$ and $m_s = 0$ states, and the Zeeman shift due to an applied magnetic field \cite{Loubser_RPP_1978}.  We applied a $\sim\!45\,\mathrm{G}$ magnetic field along the defect's symmetry axis to separate the $m_s = \pm 1$ states by $\sim\,$\unit{250}{\mega\hertz}. Consistent with measurements performed at lower temperatures \cite{Acosta_PRL_2010, Chen_APL_2011}, we observed shifts in $D$ on the order of $100\,\mathrm{kHz}/\mathrm{K}$ due to lattice expansion.  In addition to this shift, both $C$ and the off-resonant $I_\mathrm{PL}$ showed a pronounced decay above \unit{550}{\kelvin}; by \unit{700}{\kelvin} the Lorentzian dips were barely observable and the off-resonant $I_\mathrm{PL}$ dropped to $\sim\!20\%$ of its room-temperature value [Fig. \ref{fig1}(c)].

\begin{figure}
  \includegraphics[width=\columnwidth]{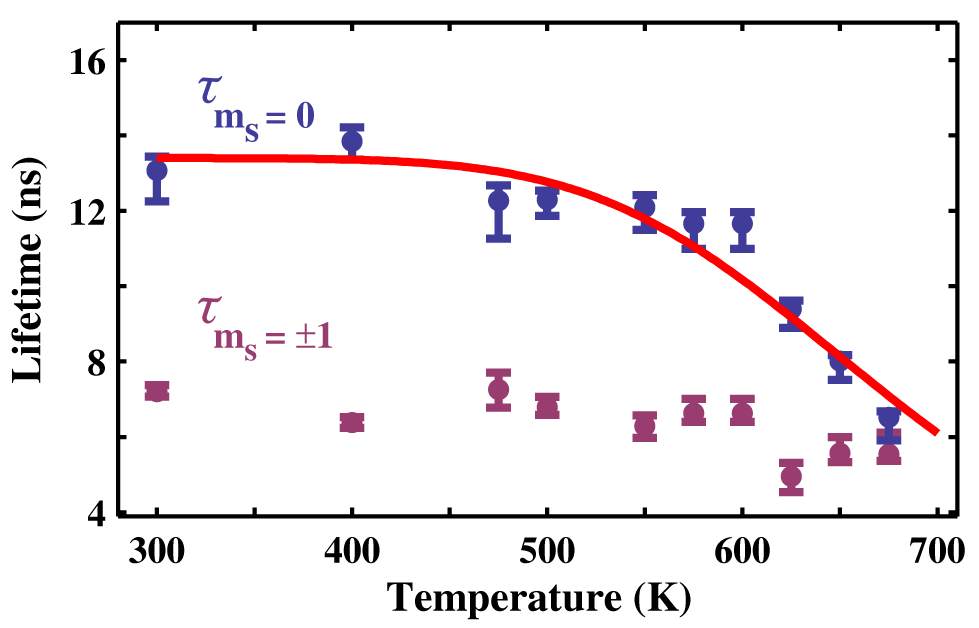}\\
  \caption{\label{figure_lifetimes}(color online) Temperature-dependent excited-state lifetimes, $\tau_{m_s = 0}\left(T\right)$ (blue) and $\tau_{m_s = \pm 1}\left(T\right)$ (magenta), for a single NV center.  The error bars reflect fitting uncertainties (68\% intervals).  The red line is the Mott-Seitz formula for $\tau_{m_s = 0}\left(T\right)$ with $\tau_{m_s = 0} \left(300\,\mathrm{K}\right) = 13.4\,\mathrm{ns}$, $s_{m_s = 0} = 3420$, and $\Delta E = 0.48\,\mathrm{eV}$.}
\end{figure}

Thermal quenching of a point defect's $I_\mathrm{PL}$ often results from thermally activated nonradiative processes that shorten the effective optical lifetime of an emitter \cite{DiBartolo}. Previous ensemble measurements of the temperature-dependence of the NV center's optical lifetime in bulk diamond \cite{Collins_JPhysC_1983} and nanodiamonds \cite{Plakhotnik_PCCP_2010} have shown conflicting results. We therefore have measured the spin-resolved excited-state lifetimes ($\tau_{m_s = 0}$ and $\tau_{m_s = \pm 1}$) for a single NV center as a function of temperature following the method of Ref. \cite{Robledo_NJP_2011} to investigate their influence on ESR contrast and $I_\mathrm{PL}$ [Fig. \ref{figure_lifetimes}].  The spin-dependence of the $^{3}E$ ISC preferentially shortens $\tau_{m_s = \pm 1}$ and thus optical excitation leads to biexponential fluorescence decay, where the decay constants correspond to $\tau_{m_s = 0}$ and $\tau_{m_s = \pm 1}$ and the relative amplitudes reflect the spin polarization. To probe the lifetimes, we initialized the spin into $m_s = 0$ using a \unit{1.4}{\micro\second} pulse from a \unit{532}{\nano\meter} laser. After waiting \unit{500}{\nano\second} for the singlets to depopulate, we applied a resonant microwave pulse to rotate the spin into a superposition of $m_s = 0$ and $m_s = -1$. We then applied a \unit{555}{\nano\meter} picosecond laser pulse and measured the resulting PL with a time-correlated photon counting module. We repeated this measurement for varying spin rotation angles and performed global Bayesian inference to infer $\tau_{m_s = 0}$, $\tau_{m_s = \pm 1}$, and the finite spin polarization at each rotation angle from the measurements (see Ref. \cite{SOM} and references \cite{VrugtSIAM, LaloyVrugt, Brett_2005, ChenShao1999,GelmanRubin1992} therein). Above \unit{550}{\kelvin}, $\tau_{m_s = 0}\left(T\right)$ showed a sharp reduction with increasing temperature.  In spite of this, we found the ground-state spin polarization after optical pumping was temperature independent within our measurement uncertainty up to \unit{650}{\kelvin} \cite{SOM}.

We find that $\tau_{m_s = 0}\left(T\right)$ is accurately described by the Mott-Seitz formula for nonradiative relaxation via multiphonon emission \cite{DiBartolo},
\begin{equation}\label{mottseitz}
\tau_{m_s = 0} \left(T\right) = \frac{\tau_{m_s = 0} \left(300\,\mathrm{K}\right)}{1 + s_{m_s = 0} \exp \left[-\frac{\Delta E}{k_{\mathrm{B}} T}\right]},
\end{equation}
where $1/\tau_{m_s = 0} \left(300\,\mathrm{K}\right) = 1/\tau_{\mathrm{rad}} + k_{m_s =0}$.  Here $1/\tau_{\mathrm{rad}}$ is the radiative rate, $k_{m_s =0}$ is the room temperature nonradiative transition rate from the $^{3}E$ $m_s = 0$ sublevel to the uppermost singlet state [Fig. \ref{fig1}(a)], $s_{m_s = 0}$ is the frequency factor, and $\Delta E$ is the energy barrier for the nonradiative process.  From a fit to $\tau_{m_s= 0} \left(T\right)$, we find $\tau_{m_s = 0} \left(300\,\mathrm{K}\right) = \left(13.4 \pm 0.6\right)\,\mathrm{ns}$, $s_{m_s = 0} < 1.64 \times 10^{4}$, and $\Delta E = \left(0.48\,^{+0.15}_{-0.13}\right)\,\mathrm{eV}$.  In general, $\Delta E$ is interpreted via a one-dimensional configuration coordinate diagram. These diagrams show the dependence of total energies in different electronic states as a function of a generalized configuration coordinate $\Delta R$. The latter measures the total displacement of all atoms along the path that interpolates between equilibrium geometries in the relevant electronic states. $\Delta E$ is then classically defined as the energy difference between the intersection point of two potential energy curves and the energy minimum of the upper curve. In such a scenario, nonradiative relaxation is interpreted as a phonon-assisted process over the energy barrier.

\begin{figure}
  \includegraphics[width=\columnwidth]{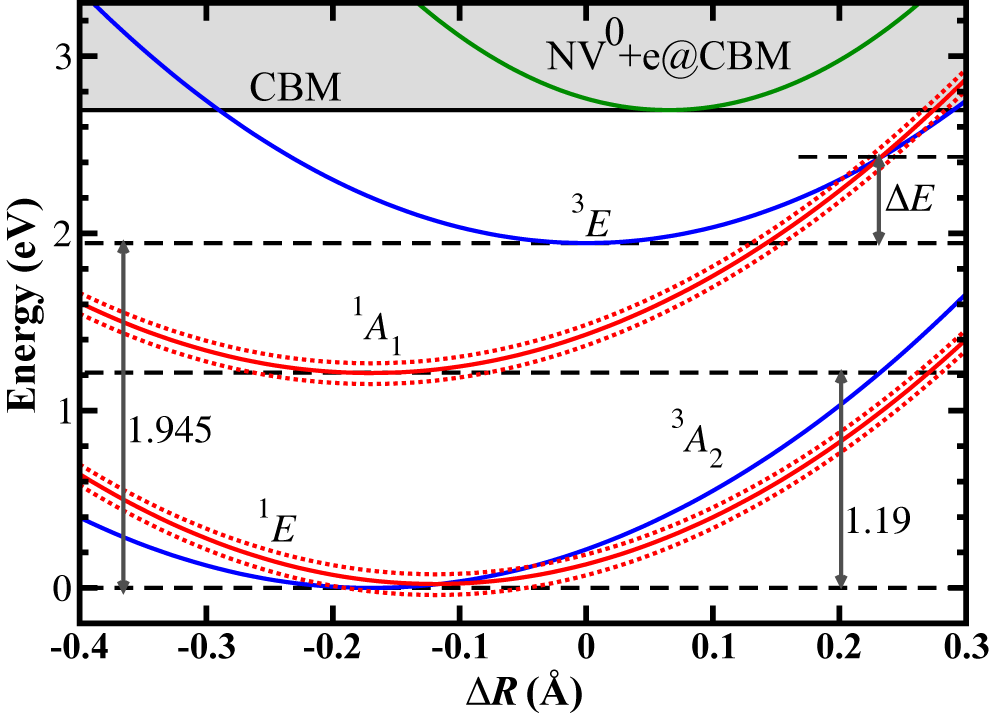}\\
  \caption{\label{audrius_fig}(color online) Configuration coordinate diagram of the NV$^{-}$ center. $\Delta R$ measures the total displacement of all atoms along the path that interpolates between equilibrium geometries in the $^{3}A_2$ and $^{3}E$ electronic states. All energies are relative to the $^{3}A_2$ state's equilibrium energy. We placed the $^{1}A_1$ state based on the assumption that its equilibrium atomic configuration and vibrational frequencies were similar to those of the $^{3}A_2$ state and using our measurement of $\Delta E$. The $^{1}E$ state was placed $1.19\,\mathrm{eV}$ below the $^{1}A_1$ state based on their known energy difference. The uncertainties in the $^{1}A_1$ and $^{1}E$ energies reflect the uncertainty in $\Delta E$ ($68\%$ interval, dashed red lines). The triplet ($1.945\,\mathrm{eV}$) and singlet ($1.19\,\mathrm{eV}$) zero phonon lines are also indicated.}
\end{figure}
We combine our measurement of $\Delta E$ with electronic structure calculations to construct a configuration coordinate diagram (CCD) for the NV$^{-}$ center in its $^{3}A_{2}$, $^{3}E$, $^{1}A_{1}$, and $^{1}E$ electronic states. The theoretical methodology employed relies on the use of supercells to describe defects and hybrid density functionals for an accurate treatment of the electronic structure \cite{Weber_PNAS_2010,Gali_PRL_2009}.  Additional details regarding the computational methods are provided in the Supplementary Material \cite{SOM} and references \cite{HSE06,Bloechl_PRB_1994, Kresse_CMS_1996, Fu_PRL_2009,Abtew_PRL_2011,Sato_PRB_2002, Ranjbar_PRB_2011, levine} therein. The calculated ZPL between the triplet states, $1.972\,\mathrm{eV}$, agrees favorably with the experimental value of $1.945\,\mathrm{eV}$.  In Fig. \ref{audrius_fig}, we placed the minima of $^{3}A_2$ and $^{3}E$ states according to the experimental ZPL with zero defined by the $^{3}A_{2}$ state's equilibrium energy. We mapped the potential energy surfaces for these two states along the path that linearly interpolates between their equilibrium geometries.  The resulting Stokes and anti-Stokes shifts were $0.27\,\mathrm{eV}$ and $0.23\,\mathrm{eV}$, respectively.

In constructing the CCD for the singlet states, we also assumed that the $^{1}A_{1}$ state's equilibrium atomic configuration and vibrational frequencies were equal to those of the $^{3}A_{2}$ state.  Indeed, the  $^{1}A_{1}$ state's $a_{1}^{2}e^{2}$ electron configuration \cite{Maze_NJP_2011, Doherty_NJP_2011} implies the two states have a similar electron density.  In turn, this implies that the ionic forces are similar for these two states away from their equilibrium geometries. Therefore, we conclude the Huang-Rhys factor ($S$), the average number of nonequilibrium phonons emitted when the electronic state changes \cite{DiBartolo}, for the $^{3}E$ to $^{1}A_1$ transition is the same as that for the $^{3}E$ to $^{3}A_{2}$ optical transition.  For the latter our calculations give a value of $S \approx 3.2$, close to the value of $S \approx 3.65$ inferred from the relative spectral weight of the ZPL with respect to the phonon sideband \cite{Davies_1976}.  The value $S > 1$ for the $^{3}E$ to $^{1}A_1$ transition confirms our hypothesis, implied in Eq. \eqref{mottseitz}, that the ISC is accompanied by multiphonon emission.

We interpret $\Delta E$ as the energy difference between the intersection point of the $^{3}E$ and $^{1}A_{1}$ potential energy curves and the energy minimum of the $^{3}E$ state. This places the $^{1}A_{1}$ state $\left(0.76 \pm 0.07\right)\,\mathrm{eV}$ below the $^{3}E$ state. By fixing $\Delta E$ based on our measurements, our analysis avoids difficulties in computing the absolute energies of the singlet states resulting from their multi-determinant electronic wavefunctions \cite{Maze_NJP_2011, Doherty_NJP_2011}. We placed the $^{1}E$ state $1.19\,\mathrm{eV}$ below the $^{1}A_{1}$ state \cite{Rogers_NJP_2008,Acosta_PRB_2010,Manson_arXiv_2010} and find that the $^{1}E$ singlet is close in energy to the $^{3}A_2$ triplet. Since the $^{3}A_2$ state is the ground state, the $^{1}E$ state should be higher in energy. Within the uncertainty in $\Delta E$, the configuration coordinate diagram fulfills this constraint and indicates that only small energy barriers for nonradiative relaxation from the $^{1}E$ state are possible in qualitative agreement with energy barriers inferred from the temperature-dependence of the $^{1}E$ state lifetime \cite{Acosta_PRB_2010, Robledo_NJP_2011}.

To demonstrate that the neutral charge state of the NV center (NV$^{0}$) does not interfere in the temperature-activated non-radiative processes discussed above, in Fig. \ref{audrius_fig} we also placed the potential energy curve for the NV$^{0}$ center in its $^{2}E$ ground state based on the charge-state transition levels given in Ref. \cite{Weber_PNAS_2010} and the calculated total-quadratic distortion $\Delta R$.  The resulting high ionization potential for the NV$^{-}$ center in the $^{3}E$ state ($>$ 0.8 eV) shows that the escape of the electron to the conduction band minimum is negligible for the temperatures studied.

Our inference of the $^{1}A_{1}$ and $^{1}E$ energies is influenced by our classical interpretation of $\Delta E$, which is based on the consideration of a single effective phonon mode of $a_1$ symmetry that strongly couples to the distortion of the defect \cite{DiBartolo}.  In fact, the temperature-activated ISC from the $^{3}E$ $m_s = 0$ state to the $^{1}A_{1}$ state also requires coupling to phonons of $e$ symmetry, while this is unnecessary for the ISC from the $^{3}E$ $m_s = \pm 1$ states \cite{Manson_arXiv_2010}.  Nevertheless, the presented model is a good approximation that provides a suitable description of the temperature-dependent ISC.  The generalization of this model that takes into account quantum effects as well as coupling to phonons of different symmetry during radiative and non-radiative processes will be presented elsewhere \cite{Alkauskas_2012}.

\begin{figure}
  \includegraphics[width=\columnwidth]{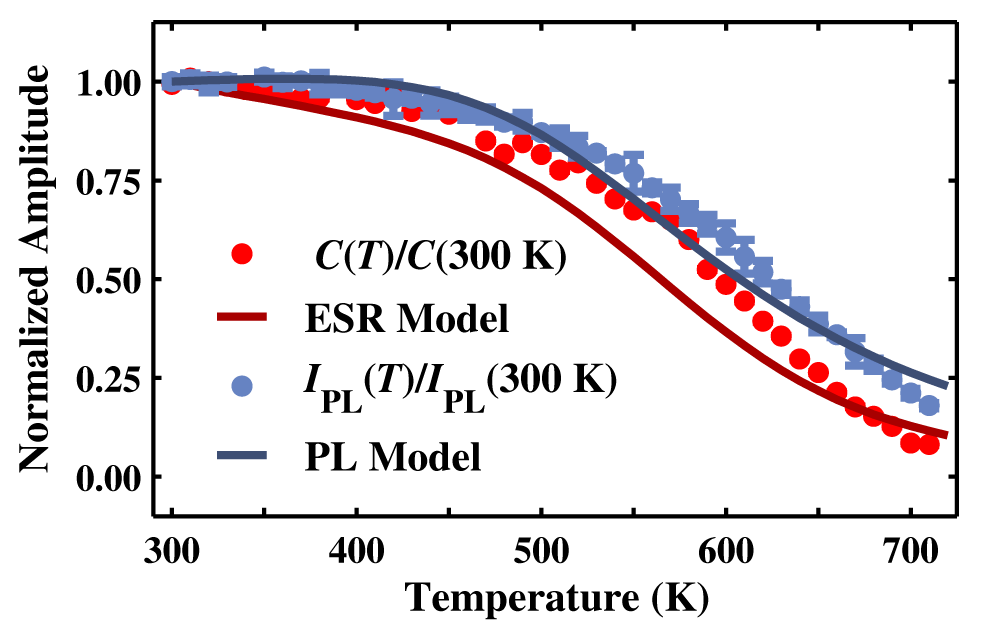}\\
  \caption{\label{figure_model}(color online) Temperature-dependent ESR contrast ($C(T)$, red points) and $I_\mathrm{PL}\left(T\right)$ (blue points).  Both quantities were normalized by their \unit{300}{\kelvin} values ($\sim$\!$0.22$ and $\sim$\!$80\,\mathrm{kCps}$, respectively). $C(T)$ was determined from fitting the CW ESR data and $I_\mathrm{PL}$ was determined from Gaussian fits to confocal microscopy line scans. The solid lines represent predictions for $C(T)/C(300\,\mathrm{K})$ and $I_\mathrm{PL}\left(T\right)/I_\mathrm{PL}\left(300\,\mathrm{K}\right)$ from a model of the CW ESR experiments depending on $\tau_{m_s = 0}\left(T\right)$ and $\tau_{m_s = \pm 1}\left(T\right)$.}
\end{figure}
Having interpreted $\Delta E$, we employ a model of the ESR experiments using the density matrix formalism that includes dissipation to reveal that the observed decreases in $C$ and $I_\mathrm{PL}$ are predominantly caused by the temperature dependence of $\tau_{m_s = 0}$ and $\tau_{m_s = \pm 1}$~\cite{Rand_2010}. The model consists of irreversible transitions between five effective energy levels [Fig. \ref{fig1}(a)] under optical pumping with an off-diagonal term to capture the coherent microwave driving \cite{SOM}. The only temperature-dependent components are $\tau_{m_s = 0}$, $\tau_{m_s = \pm 1}$, and the known temperature-dependence of the effective singlet lifetime \cite{Robledo_NJP_2011}. The model shows agreement with our measurements of $C(T)/C(300\,\mathrm{K})$ and $I_{\mathrm{PL}}(T)/I_{\mathrm{PL}}(300\,\mathrm{K})$ [Fig.~\ref{figure_model}].  It reveals that the small decrease in $C$ below \unit{550}{\kelvin} is due to shortening of the singlet lifetime while the decreases in both $C$ and $I_\mathrm{PL}$ above \unit{550}{\kelvin} are primarily caused by the thermally-activated nonradiative processes that shorten $\tau_{m_s = 0}$ and $\tau_{m_s = \pm 1}$.

\begin{figure}[t]
  \includegraphics[width=\columnwidth]{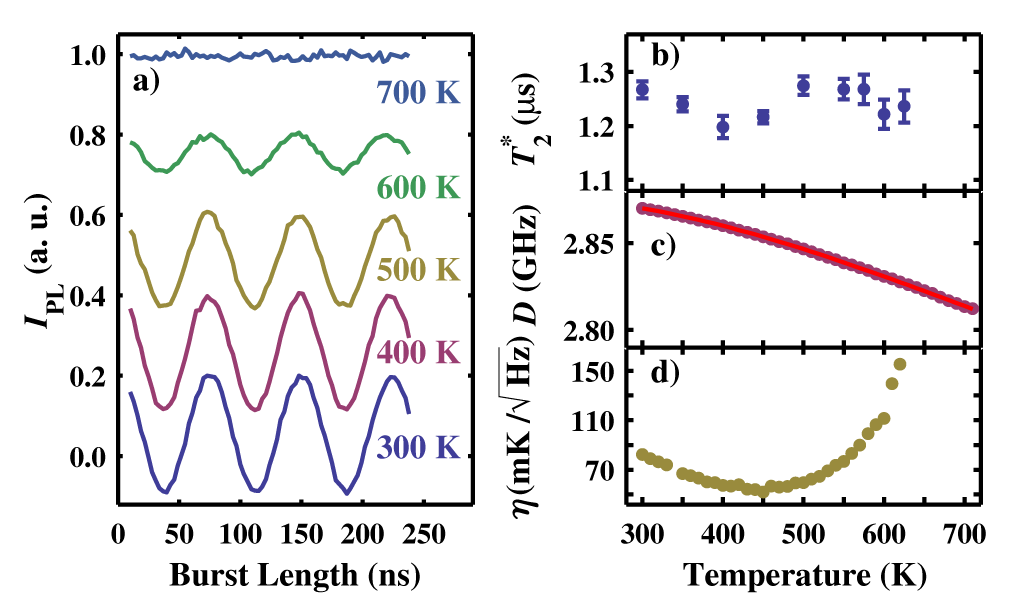}\\
  \caption{\label{figure5}(color online) (a) Rabi oscillations on the $m_s = 0$ to $m_s = -1$ transition at varying temperatures. The measurements are offset for clarity. (b) $T_2^{*}$ versus temperature inferred from the decay envelope of fits to Ramsey oscillations. These measurements were performed at $\sim$\!500\,G to polarize the NV center's $^{14}\mathrm{N}$ nuclear spin to simplify its hyperfine spectrum.  All other measurements were performed at $\sim\!45\,\mathrm{G}$. Error bars represent $68\%$ intervals. (c) The crystal field splitting, $D\left(T\right)$, inferred from Lorentzian fits to the CW ESR data.  The solid red line is the third-order polynomial fit to the data given in the main text. (d) Single-spin thermal sensitivity ($\eta$). $\eta$ was calculated using Eq. \eqref{thermalsens} with all parameters from the same NV center.}
\end{figure}

We established from pulsed ESR measurements that the NV center remains spin coherent even as nonradiative processes diminish the ESR contrast and $I_\mathrm{PL}$. The spin's robust coherence is due to the equal population of the sublevels of the $^{13}\mathrm{C}$ nuclear spin impurities which decohere the spin at all but cryogenic temperatures \cite{Childress_Science_2006}. We performed these measurements by polarizing the spin with a \unit{532}{\nano\meter} CW laser pulse, coherently manipulating the spin with resonant microwaves, and reading out the spin state by monitoring the PL \cite{Jelezko_PRL_2004}. Both Rabi [Fig. \ref{figure5}(a)] and Ramsey pulse sequences were employed (see Ref. \cite{SOM} and references \cite{Jacques_PRL_2009, Smeltzer_NJP_2011} therein), demonstrating that the spin can be coherently controlled at temperatures exceeding \unit{600}{\kelvin} and showing $I_\mathrm{PL}$ and ESR contrast trends consistent with the fluorescence quenching observed in our CW ESR measurements. Moreover, fits to the Ramsey oscillation decay envelope showed $T_2^{*}$ is temperature independent up to at least \unit{625}{\kelvin} [Fig. \ref{figure5}(b)].  We also performed longitudinal relaxation measurements by optically polarizing the spin into $m_s = 0$ and measuring the relaxation into $m_s = \pm 1$ through the time-dependent $I_\mathrm{PL}$.  The measurements demonstrated that $T_1$ at \unit{600}{\kelvin} (\unit{340 \pm 50}{\micro\second}) was still much longer than the period of the pulsed ESR measurements, confirming that spin relaxation does not explain the observed decrease in the ESR contrast and $I_\mathrm{PL}$.

The persistence of the spin coherence and the strong temperature dependence of the crystal field splitting, $D$, suggest the possibility of using the spin resonances for thermometry. Based on our CW ESR measurements, we find that $D(T)$ [Fig. \ref{figure5}(c)] is accurately described by a third-order polynomial, $D(T) = a_0 + a_1T + a_2T^2 + a_3T^3$, between \unit{300}{\kelvin} and \unit{700}{\kelvin}, with $a_0 = \left(2.8697 \pm
0.0009\right)\,\mathrm{GHz}$, $a_1 = \left(9.7 \pm 0.6\right) \times
10^{-5}\,\mathrm{GHz}/\mathrm{K}$, $a_2 = \left(-3.7 \pm 0.1\right) \times
10^{-7}\,\mathrm{GHz}/\mathrm{K}^{2}$, and $a_3 = \left( 1.7 \pm 0.1\right)
\times 10^{-10}\,\mathrm{GHz}/\mathrm{K}^{3}$.  The derivative of this polynomial yields thermal shifts ranging from \unit{80}{\kilo\hertz\per\kelvin} at \unit{300}{\kelvin} to \unit{170}{\kilo\hertz\per\kelvin} at \unit{700}{\kelvin}.  The possibility of spin-based thermometry is further motivated by the established methods for measuring small static shifts in the NV center spin resonance frequencies \cite{Maze_Nature_2008,Balasubramanian_Nature_2008,Acosta_APL_2010,Schoenfeld_PRL_2011}.  Perhaps the best known method for sensing such shifts is through a Ramsey measurement, in which a superposition of two spin states is created and evolves under the influence of static dephasing fields.  After a time $t$, a microwave control pulse is applied to convert the accumulated phase into a population difference of the two spin states that can be read out optically.

We quantify the thermal sensitivity ($\eta$) based on our measurements of the defects's temperature-dependent optical and spin properties using an expression analogous to those derived for NV center DC magnetic field sensing with a Ramsey pulse sequence \cite{Taylor_NatPhys_2008}:
\begin{equation}\label{thermalsens}
\eta\left(T\right) = \frac{1}{2\pi \frac{dD\left(T\right)}{dT} A\left(T\right) \sqrt{T_2^{*}}}.
\end{equation}
 $A(T)$ accounts for the finite photon count rate and ESR contrast and is given by
 \begin{equation}\label{A}
 A\left(T\right) = {\left(1+2\frac{\alpha_0+\alpha_1}{\left(\alpha_0-\alpha_1\right)^2}\right)}^{-\frac{1}{2}},
 \end{equation}
where $\alpha_0 = \tau_\mathrm{F} I_\mathrm{PL}\left(T\right)$ and $\alpha_1
= \tau_\mathrm{F}
I_\mathrm{PL}\left(T\right)\left(1-C_\mathrm{P}\left(T\right)\right)$. Here
$\tau_\mathrm{F}$ is the fluorescence readout duration (\unit{350}{\nano\second}) and
$C_\mathrm{P}\left(T\right)$ is the contrast of a pulsed ESR measurement (0.24 for our Ramsey measurements).  Based on the data presented in Fig. \ref{figure_model} and Fig. \ref{figure5}(b,c), we find that $A\left(300\,\mathrm{K}\right) \sim 0.02$ and $\eta$ $\sim 100\,\mathrm{mK}/\sqrt{\mathrm{Hz}}$ between room temperature and \unit{600}{\kelvin} [Fig. \ref{figure5}(d)].  At higher temperatures, the sensitivity becomes limited by the quenching of the optical spin readout. This sensitivity could be enhanced for NV centers in isotopically pure diamond, which can exhibit $T_2^{*}$ times greater than \unit{100}{\micro\second} \cite{Zhao_arXiv_2012}, corresponding to $\eta$ better than $10\,\mathrm{mK}/\sqrt{\mathrm{Hz}}$.

These results demonstrate that NV centers offer operating temperatures and thermal sensitivities relevant to studies of thermal gradients in a variety of microscale and nanoscale systems.  Recent demonstrations of the incorporation and quantum control of nanocrystal diamonds containing NV centers in living cells \cite{McGuinness_NatNano_2011} suggest that NV centers could provide an alternative to quantum dot nanostructures for cellular thermometry \cite{Yang_ACSNano_2011}.  In addition, as the ideal thermal and mechanical properties of diamond have motivated its use in scanning thermal microscopy tips \cite{Majumdar_RevSciInst_1995}, advances in the incorporation of NV centers in scanning probe tips \cite{Maletinsky_NatNano_2012, Rondin_APL_2012} indicate the possibility of spin-based thermal microscopy for investigating materials such as nanostructured thermoelectrics \cite{Soini_APL_2010, Shi_JHT_2003}. Finally, given the large thermal shifts in $D$ and the ability to optically polarize the spin at high temperatures, high-density ensembles of NV centers \cite{Acosta_PRB_2009} could be used as an in-situ thermometer for electron paramagnetic resonance experiments in analogy with chemical-shift thermometers used in nuclear magnetic resonance \cite{Bielecki_JMagR_1995}.

The results presented here provide fundamental insights on how thermally-driven nonradiative processes can influence the fluorescence properties of defect-based spin qubits.  We show that the NV center's fluorescence-based spin readout is ultimately limited above room temperature by nonradiative processes that diminish the spin-selectivity of its $^{3}E$ to $^{1}A_1$ orbital transition and upset the balance between the defect's radiative and nonradiative relaxation rates. Coupled with the persistence of its spin coherence, these measurements suggest the NV center could find application as a nanoscale thermal sensor over a broad temperature range.

This work was funded by the AFOSR, ARO, and DARPA. We acknowledge computational resources at XSEDE. A portion of this work was done in the UCSB nanofabrication facility, part of the NSF funded NNIN network. AA was supported by a grant from the Swiss NSF.


%

\end{document}


\title{Supplemental Material for \\
``Measurement and Control of Single Nitrogen-Vacancy Center Spins above 600\,K''}
\author{D. M. Toyli}
 \thanks{These two authors contributed equally to this work.}
 \affiliation{Center for Spintronics and Quantum Computation,\\ University of
California, Santa Barbara, CA 93106-6105, USA}
\author{D. J. Christle}
 \thanks{These two authors contributed equally to this work.}
 \affiliation{Center for Spintronics and Quantum Computation,\\ University of
California, Santa Barbara, CA 93106-6105, USA}
\author{A. Alkauskas}
\affiliation{Center for Spintronics and Quantum Computation,\\ University of
California, Santa Barbara, CA 93106-6105, USA} \affiliation{Materials
Department,\\ University of California, Santa Barbara, CA 93106-5050, USA}

\author{B.\ B.\ Buckley}
\affiliation{Center for Spintronics and Quantum Computation,\\ University of
California, Santa Barbara, CA 93106-6105, USA}
\author{C. G. Van de Walle}
\affiliation{Center for Spintronics and Quantum Computation,\\ University of
California, Santa Barbara, CA 93106-6105, USA} \affiliation{Materials
Department,\\ University of California, Santa Barbara, CA 93106-5050, USA}
\author{D.\ D.\ Awschalom}
\email[Corresponding author.\\ Email address: ]{awsch@physics.ucsb.edu}
\affiliation{Center for Spintronics and Quantum Computation,\\ University of
California, Santa Barbara, CA 93106-6105, USA}


\date{\today}
\maketitle

\SIunits[thinspace,thinqspace]


\renewcommand{\thesection}{S.\arabic{section}}
\renewcommand{\thesubsection}{\thesection.\arabic{subsection}}

%
\makeatletter 
\def\tagform@#1{\maketag@@@{(S\ignorespaces#1\unskip\@@italiccorr)}}
\makeatother

\makeatletter \makeatletter \renewcommand{\fnum@figure}
{\figurename~\thefigure} \makeatother

\section{Methods and supplementary data}

\subsection{Sample fabrication}

The sample used in this work was a \unit{4}{\milli\meter} x
\unit{4}{\milli\meter} x \unit{0.5}{\milli\meter} (100) Electronic Grade CVD
diamond from Element Six. The sample was irradiated with $2 \times
10^{14}/\mathrm{cm}^2$, $2\,\mathrm{MeV}$ electrons and annealed at
\unit{800}{\celsius} in forming gas to increase the nitrogen-vacancy (NV)
center density. The solid immersion lenses were etched using a Ga$^{+}$
focused ion beam system (FEI Helios 600) at an accelerating voltage of
\unit{30}{\kilo\volt} and an etch current of \unit{2.8}{\nano\ampere}. After
milling, we used an Ar/Cl$_{2}$ inductively-coupled plasma etch to remove
approximately \unit{80}{\nano\meter} from the surface to remove damage from
the Ga$^{+}$ ion implantation. A thick Ti/Pt metallization was used for the
heaters and ring antennas (\unit{10}{\nano\meter} Ti,
\unit{400}{\nano\meter} Pt) so that the devices would operate well in air at
elevated temperatures. The area of the diamond where the resistive
temperature detector (Omega Engineering, RTD-3-F3105-72-G-B) was bonded with
ceramic adhesive (Cotronics, Durabond 950) was first roughened with an
oxygen plasma to promote adhesion. The samples were mounted with ceramic
adhesive on glass to thermally isolate the diamond and were packaged in
brass holders with coplanar waveguide transmission lines for microwave
connections. The temperature of the sample was controlled with a PID
feedback loop using a Lakeshore 340 temperature controller. The temperature
stability of the sample was measured using the Lakeshore controller to be
within $\pm 50\,\mathrm{mK}$ of the temperature setpoint at all temperatures
investigated. We also independently calibrated our thermometry by heating
the sample with the controller feedback loop and separately measuring the
sample temperature with an infrared microscope and found good agreement
between the two measurements.

\subsection{Experimental setup}
The confocal microscope setup used for this work was similar to the
microscope used for angle-resolved magneto-photoluminescence studies
described in Ref. \cite{Epstein_NatPhys_2005}. Here we will describe
additional experimental details relevant to the work described in the main
text.

A 0.7\,NA microscope objective (Mitutoyo Plan Apo) with a
\unit{6.5}{\milli\meter} working distance was used in order to provide
clearance for the RTD on the sample surface. The overall timing of the pulsed
measurements was controlled by both a pulse pattern generator (Agilent,
81110A) and a digital delay generator (Stanford Research Systems, DG645).
Pulsed microwaves for ESR measurements were generated with a signal
generator, I/Q modulator, and arbitrary waveform generator from National
Instruments (PXIe-5673E) in series with a microwave amplifier (Amplifier
Research, 5S1G4). An acousto-optic modulator (AOM) from Isomet (1250C) was
used to modulate the \unit{532}{\nano\meter} CW laser (Oxxius,
532S-150-COL-OE) for pulsed ESR measurements. The voltage output from the AOM
driver (Isomet, 525C-2) was optimized with a microwave switch, attenuators,
and amplifier to provide an optical extinction of $\sim\!50\,\mathrm{dB}$ in
a single pass configuration. Photon counting was performed with a silicon
avalanche photodiode (Perkin Elmer, SPCM-AQRH-15-FC).  For all measurements
discussed in the main text the NV center's fluorescence was collected between
the wavelengths of \unit{630}{\nano\meter} and \unit{800}{\nano\meter}.  It
has been shown previously that the NV center's phonon sideband does not
significantly broaden over the range of temperatures measured in this work
\cite{Plakhotnik_PCCP_2010}.

Measurements of the optical lifetime were carried out using a mode-locked
Ti:Sapphire laser (Coherent Mira 900) pumped by a \unit{7}{\watt}
\unit{532}{\nano\meter} Verdi G7 laser. The infrared output of approximately
\unit{1}{\watt} is fed into an optical parametric oscillator (Coherent
Mira-OPO) tuned to output at \unit{555}{\nano\meter}. We use a
\unit{20}{\meter} long multimode fiber (OZ Optics,
QMMJ-33-UVVIS-10/125-3A-20) to increase the pulse length to
$\sim\!5\,\mathrm{ps}$. An electro-optic modulator (Conoptics 350) and
pulse-picking electronics were used to reduce the \unit{76}{\mega\hertz}
repetition rate of the \unit{555}{\nano\meter} pulse train. A time-correlated
photon counting module (PicoHarp 300) measured the arrival time of photons
into the avalanche photodiode relative to the pulse-picked
\unit{555}{\nano\meter} light pulses.

\subsection{NV center optical lifetime fits}

We applied a Bayesian approach to infer the spin-resolved fluorescence
lifetime parameters and their associated intervals for the uncertainties
shown in the main text. The difficulty of fitting sums of exponentials to
experimental data is a known problem and originates from the high
correlation between amplitude and lifetime parameters that tends to confound
them \cite{Brett_2005}. Our approach forgoes the use of asymptotic
approximations in favor of calculating the actual joint posterior density to
more accurately infer the parameters and their associated errors.

Before we analyze the data, we remove approximately \unit{2.5}{\nano\second}
of histogram data after the initial onset of fluorescence to eliminate
fast-decaying background photoluminescence detected in the sample. Because
of the finite extinction of our electro-optical modulator, very small
$\sim\!$2 \,ns fluorescence bursts with a \unit{76}{\mega\hertz} spacing
appear in our raw data. Of these, we removed the first two and truncated the
dataset by removing points including and after the third unwanted pulse. At
each temperature, we found about six spin rotation angles were needed for
good resolution of the separate lifetimes. At the driving strength used for
these measurements, we could perform a $\pi$ rotation in
\unit{80}{\nano\second}.

For our analysis, we assume that for each microwave burst length, the
fluorescence decay follows biexponential decay with a constant offset subject
to random error,

\begin{equation}
f_{ij}\left(t_k\right) = A_j  \exp\left(-t_k/\tau_{m_s=0,i}\right) + B_j\exp\left(-t_k\left(1+k_i\right)/\tau_{m_s = 0,i}\right) + C_j + \epsilon,
\end{equation}
where $\epsilon$ is normally distributed with mean $0$ and variance $\eta^2
\sigma_\mathrm{Poisson}^2$. Here $\sigma_\mathrm{Poisson}$ is the square root
of the noiseless value of $f_{ij}\left(t_k\right)$ and $\eta$ is a
dimensionless scaling parameter strictly above unity that accounts for other
random noise in the experiment. This value is inferred from analysis of
experimental data to be very close to unity, serving as a measure of goodness
of fit that indicates our measurements are nearly shot-noise limited, and
that any systematic effects from unextinguished pulses are within the random
error of the measurement. For each microwave burst length, $A_j$, $B_j$, and
$C_j$ are free parameters while parameters $\tau_{m_s=0,i}$ and $k_i$ are
shared between burst lengths at a given temperature. $\tau_{m_s=\pm 1,i}$ is
related to $\tau_{m_s=0,i}$ through the rate relation,
\begin{equation}
\frac{1}{\tau_{m_s=\pm 1, i}} = \frac{1}{\tau_{m_s = 0, i}} + k_i,
\end{equation}
where $k_i$ is the difference in excited state intersystem crossing rates
between the two spin states. All parameters in the analysis use
non-informative and proper (bounded) prior distributions and all posterior
distributions were found to be relatively insensitive to the particular form
of the prior.

Because the lifetimes are shared globally between data sets at the same
temperature, the posterior density typically exceeds 30 dimensions, making
the use of standard numerical integration techniques infeasible. To compute
marginal posterior densities for the lifetimes plotted in the main text, we
employ the DREAM$_{\mathrm{ZS}}$ algorithm written in MATLAB
\cite{VrugtSIAM,LaloyVrugt}, a Markov Chain Monte Carlo (MCMC) approach that
uses adaptive proposal distribution tuning, sampling from the past, and
snooker update on parallel chains to rapidly explore high-dimensional
posterior distributions. In MCMC, each of the $N$ chains executes a random
walk through the parameter space following a modified Metropolis-Hastings
rule to control whether a proposed $d$-dimensional move is accepted or
rejected. Because the algorithm maintains detailed balance at each step, the
target distribution after a burn-in period is the desired posterior
distribution. We find good results using the recommended settings along with
$N = 8$ parallel chains. Convergence to the target distribution was assessed
both graphically and with the Gelman-Rubin statistic, $\hat{R} < 1.05$
\cite{GelmanRubin1992}. The point estimates for the lifetimes and other
parameters are computed from the respective sample empirical means, and the
highest posterior density intervals are computed using the method of Chen and
Shao \cite{ChenShao1999}.

To infer the parameters in the Mott-Seitz equation for non-radiative decay,
we apply a similar technique of bounding the base lifetime, frequency
factor, and energy barrier and computing the respective marginal posterior
densities. As an approximation, we assume the likelihoods for the $m_s = 0$
lifetime at each temperature are approximately normal, setting the variance
parameter equal to the empirical variance and the location parameter equal
to the empirical mean of the samples generated from the individual marginal
distributions via MCMC. This avoids the 337-dimensional exact approach,
which we found intractable even with DREAM$_{\mathrm{ZS}}$. The base
lifetime and energy barriers use uniform uninformative priors while the
frequency factor is bounded by its theoretical maximum
($\sim\!74\,\mathrm{meV}$) using a $\propto 1/s$ density to express great
uncertainty in the scale of this value; there exists some sensitivity
(within the quoted error) of the posterior on the choice of this prior since
we were unable to measure above $675\,\mathrm{K}$. The analysis also allows
for a linear systematic temperature offset up to a maximum of
$10\,\mathrm{K}$ at $700\,\mathrm{K}$, although this correction is
essentially negligible. We compute the marginal densities for the Mott-Seitz
parameters and quote summary statistics in the main text as previously
described.
\subsection{Measurements of multiple NV centers}

All measurements discussed in the main text were taken on single NV centers.
The data shown in Fig. 1, Fig. 2, and Fig. 5(b,c,d) was taken on an NV
center (NV1) in a solid immersion lens while the data in Fig. 5(a) was taken
on another NV center (NV2) in bulk diamond.  The two NV centers showed
nearly identical electron spin resonance (ESR) contrast temperature
dependencies [Fig. S1]. NV1 and NV2 showed similar spin-dependent
excited-state lifetime temperature dependencies up to \unit{625}{\kelvin},
although the lower photon collection efficiency on NV2 precluded
spin-resolved lifetime measurements on this NV center at temperatures above
\unit{625}{\kelvin}.
\begin{center}
\begin{figure}[h]
  \includegraphics[width=0.40\textwidth]{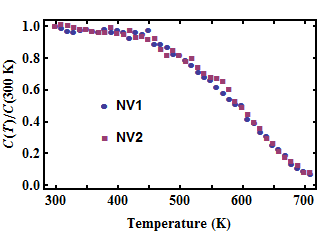}\\
  \caption{Electron spin resonance (ESR) contrast ($C(T)$) determined from
  continuous-wave (CW) ESR measurements normalized by $C(300\,\mathrm{K})$
  for two different NV centers.}\label{Fig1SOM}
\end{figure}
\end{center}

\subsection{Rate equation modeling}
\begin{figure}[h]
  \includegraphics[width=0.40\textwidth]{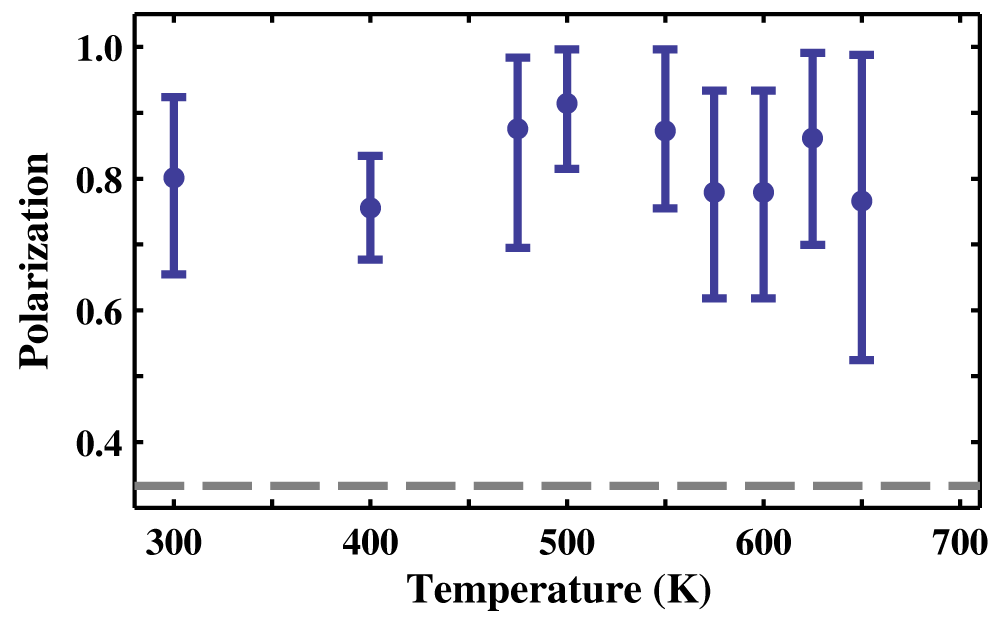}\\
  \caption{Plot of maximum spin polarization in the $m_s = 0$ spin state inferred from fits to the data with no microwaves applied (95\% intervals). The data are consistent with robust spin preparation up to at least $\sim\!650\,\mathrm{K}$. The dashed line indicates the polarization for an unpolarized spin.}\label{fig_maxpol}
\end{figure}

A simplified model of the NV center with five energy levels is used to model
the optically detected ESR signal observed in experiment
\cite{Manson_PRB_2006}. We augment the classical rate equation model of Ref.
\cite{Manson_PRB_2006} with the off-diagonal terms from the density matrix
that give rise to coherent driving in the orbital ground state with
dissipation. We numerically integrate this expression, a master equation of
the Lindblad form, to simulate the time evolution of the system under both
optical and microwave excitation that correspond to experimental conditions
\cite{Rand_2010}. We use the optical decay rates measured in our experiment
and the non-spin-conserving optical transition rates from Ref.
\cite{Manson_PRB_2006}, and assume that they are temperature-independent. To
capture the temperature dependence of the transition rates, we use the best
fit equation versus temperature for the singlet rates from Ref.
\cite{Robledo_NJP_2011} and the Mott-Seitz equation plotted in the main text
added to the room temperature intersystem crossing rates equally for the
excited spin states. The $s$ parameter in the Mott-Seitz equation was
obtained using least squares while fixing the energy and $300\,\mathrm{K}$
lifetime at the respective means of their marginal distributions. The ESR
contrast is computed by comparing the steady-state luminescence when
microwaves are applied continuously versus the luminescence without
microwaves, corresponding to the highly-detuned case we use for
normalization.  To compute the theoretical dependence of the
photoluminescence on temperature, we compute the value of the steady-state
luminescence under constant optical pumping with no applied microwaves. Both
the theoretical contrast and photoluminescence values are normalized to
unity at room temperature and plotted with experimental data to generate the
figures in the main text. The model also predicts relatively robust spin
preparation into the $m_s = 0$ sublevel in the ground state after repeated
optical cycling, which is consistent with the polarizations inferred in our
analysis shown in Fig. \ref{fig_maxpol} to within the statistical
uncertainty. We summarize the rates used in Table \ref{ratestable}.

\begin{center}
\begin{table}[h!]

\begin{tabular}{|l|l|}

  \hline
  Transition Rate & Value \\
  \hline
  $k_{42}$, $k_{31}$ & \unit{75}{\mega\hertz} \cite{Manson_PRB_2006} \\
  $k_{32}$, $k_{41}$ & \unit{1.5}{\mega\hertz} \cite{Manson_PRB_2006}\\
  $k_{51}$ & $1/\left(\left(1 + 1/1.6\right) 371\times(1 - e^{-16.6\,\mathrm{meV}/(k_B T)})\right)\,\mathrm{GHz}$ \cite{Robledo_NJP_2011} \\
  $k_{52}$ & $1/\left(\left(1 + 1.6\right) 371\times(1 - e^{-16.6\,\mathrm{meV}/(k_B T)})\right)\,\mathrm{GHz}$ \cite{Robledo_NJP_2011}\\
  $k_{45}$  & $60 + 2.552 \times 10^{5} \times e^{-0.48\,\mathrm{eV}/k_B T}\,\mathrm{MHz}$ \\
  $k_{35}$ & $2.552 \times 10^{5} \times e^{-0.48\,\mathrm{eV}/k_B T}\,\mathrm{MHz}$ \\
  $k_\mathrm{pump}$ & $0.3 \times 75\,\mathrm{MHz}$ \\
  \hline
\end{tabular}
  \caption{A summary of the numerical values used for the rate-equation
  components of the density matrix formalism.} \label{ratestable}
\end{table}
\end{center}

\subsection{Ramsey measurements}
\begin{figure}
  \includegraphics[width=0.40\textwidth]{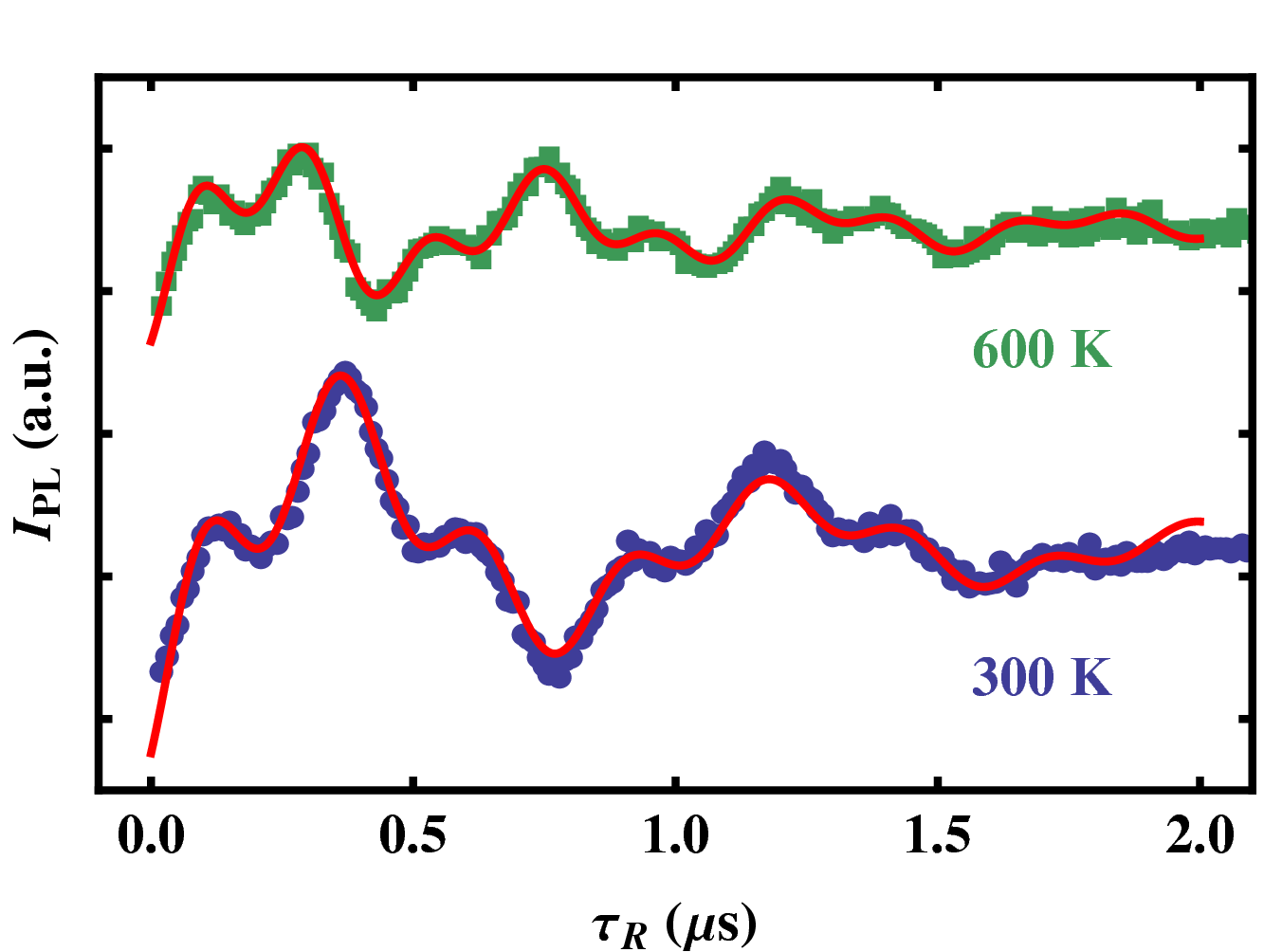}\\
  \caption{Ramsey oscillation measurements performed on NV1 at \unit{300}{\kelvin} (blue)
  and \unit{600}{\kelvin} (green).  The measurements were done at $500\,\mathrm{G}$.  The $\pi/2$ pulses
  were detuned from the lower-frequency hyperfine resonance by $-2\,\mathrm{MHz}$ on the
  $m_s=0$ to $m_s=-1$ spin transition.  The solid red lines are best fits to the
  data using Eq. S3.  The relative amplitudes of the Ramsey oscillations
  reflect the relative ESR contrasts of the two measurements.}\label{FigS4}
\end{figure}

The $T_2^{*}$ values in the main text were inferred from fits to the decay
of Ramsey oscillation measurements.  The Ramsey measurements were carried
out by first optically initializing the spin using a \unit{2}{\micro\second}
\unit{532}{\nano\meter} laser pulse. After waiting
$\sim$\unit{500}{\nano\second} for the singlet pathway to depopulate, we
rotated the spin into a superposition of its $m_s = 0$ and $m_s = -1$ spin
states with a $\pi/2$ pulse and allowed the spin to evolve for a variable
time $\tau_R$. A second $\pi/2$ pulse was used to convert the phase of the
spin acquired during free evolution into a population difference and the
spin was then read out by collecting its fluorescence for the first
$\sim\!350\,\mathrm{ns}$ of the following laser pulse. The primary NV center
studied in this work showed a hyperfine coupling of
$\sim$\unit{2.5}{\mega\hertz} to a nearby $^{13}\mathrm{C}$ ($I =
\frac{1}{2}$) in addition to its hyperfine coupling to the $^{14}\mathrm{N}$
nuclear spin ($I = 1$). To simplify the hyperfine spectrum to allow for more
accurate fits to $T_2^{*}$ we performed the Ramsey measurements at
$500\,\mathrm{G}$ to polarize the nitrogen nuclear spin into its $m_I = +1$
state \cite{Jacques_PRL_2009}. This resulted in a hyperfine spectrum with
two visible ESR transitions showing a relative polarization consistent with
Ref. \cite{Smeltzer_NJP_2011}. For the Ramsey measurements, we detuned the
$\pi/2$ pulses by -\unit{2}{\mega\hertz} from the lower frequency ESR
transition to induce oscillations in the measured PL intensity as a function
of $\tau_R$. We fit the resulting data [Fig. \ref{FigS4}] with the function
\begin{equation}
y = \exp\left(-\left(\frac{t}{T_2^{*}}\right)^2 \right)\left(A_1 \cos\left(\omega_1 t + \phi_1\right) + A_2 \cos\left(\omega_2 t + \phi_2\right)\right) + y_0
\end{equation}
to account for the two visible ESR transitions and to extract $T_2^{*}$. The
Rabi measurements shown in the main text were carried out with a similar
sampling scheme except that resonant microwave pulses of varying duration
(burst length) were applied instead of the $\pi/2$ - $\tau_R$ - $\pi/2$
portion of the Ramsey pulse sequence.

\section{Electronic structure calculations}

The configuration coordinate diagram (CCD) presented in Fig. 3 of the main
text depends on the values of Stokes (S) and anti-Stokes (AS) shifts
pertaining to the optical transition between triplet states (i.e.\
relaxation energies in the excited and ground states). While these
quantities influence greatly the shape of optical absorption and
luminescence bands, it is not straightforward to extract them from
measurements, such as those presented in Ref. \cite{Davies_1976}. Therefore,
to determine the values of these shifts, we have utilized electronic
structure calculations \cite{Gali_PRL_2009,Weber_PNAS_2010}.

The calculations in this work were performed employing the hybrid density
functional of Heyd, Scuseria, and Ernzerhnof (HSE06) to describe the NV
center's electronic structure \cite{HSE06}. Atomic cores were treated via the
projector-augmented-wave (PAW) scheme \cite{Bloechl_PRB_1994}, and the plane
wave cutoff for valence wave functions was set to 400\,eV. To simulate the
NV$^{-}$ center, we used a cubic supercell nominally containing 216 atoms.
The $\Gamma$-point was used for the Brillouin zone integration. To simulate
the excited state triplet, we used a method outlined in Ref.
\cite{Gali_PRL_2009}, whereby an electron in the spin-minority channel is
promoted to the $e$ state thus leaving a hole in the $a_1$ state. We employed
the code \verb"VASP" \cite{Kresse_CMS_1996}. A similar computational
methodology was also used in Refs. \cite{Gali_PRL_2009,Weber_PNAS_2010}.

The equilibrium geometry of the defect in the $^{3}A_2$ state has been
determined by fully relaxing the geometry. The resulting point symmetry was
$C_{3v}$, and the electron configuration $a_{1}^2e^2$. In the $^3E$ state the
$C_{3v}$ symmetry is broken, but the energy associated with symmetry breaking
is small ($\sim$0.025 eV) \cite{Abtew_PRL_2011}. In the main text it is
assumed that the $C_{3v}$ symmetry is maintained for equilibrium geometries
of all electronic states of the NV$^{-}$ center. In particular, it implies
that the symmetry is maintained along the path that interpolates between
equilibrium geometries of the defect in $^3A_2$ and $^3E$ electronic states
[cf.\ Fig.\ 3]. This reasoning neglects the occurrence of the (dynamic)
Jahn-Teller effect in the $^3E$ (and possibly $^1E$) state (see Refs.
\cite{Abtew_PRL_2011,Fu_PRL_2009} ).

The calculated parameters of the defect depend somewhat sensitively on the
adopted lattice constant. The theoretical lattice constant of diamond is
3.539 \AA, and is thus slightly underestimated with respect to the
experimental value of 3.567 \AA\ \cite{Sato_PRB_2002}. Employing the
theoretical lattice constant, we obtain the value of 2.035 eV for the zero
phonon line (ZPL) of the optical transition between the triplet $^3E$ and
$^3A_2$ states. The resulting S and AS shifts are 0.291 and 0.245 eV. If the
experimental lattice constant is adopted, the ZPL reduces to 1.972 eV (thus
in better agreement with the experimental value of 1.945 eV). Similarly, S
and AS shifts decrease, to 0.274 and 0.230 eV, respectively. Thus, for a
$\le$1 \% change in the lattice constant, the resulting change in S and AS is
more than 5\%. This is a likely consequence of the high rigidity of the
diamond lattice. We have used the parameters from the second set of
calculations to construct the CCD presented in Fig. 3. We estimate that the
uncertainty 0.02 eV in the values of the S and AS shifts translates into an
uncertainity of about 0.1 eV in the resulting position of the $^1A_1$ and
$^1E$ states. This uncertainty should be considered in addition to
experimental error bars shown in Fig. 3.

\subsection{Electron density of the $^1A_1$ state} The construction of the
configuration coordinate diagram in Fig. 3 is based on the assumption that
the equilibrium electron density of the $^1A_1$ state is very similar to that
of the $^3A_2$ state along the generalized configuration coordinate. In the
molecular orbital picture, the wavefunction of the $^3A_2$ state (spin
projection $m_{\text{s}}=+1$) is \cite{Doherty_NJP_2011,Maze_NJP_2011}:
\begin{equation}
\left | ^3A_2,m_{\text{s}}=+1 \right \rangle = \left| a_{1}\bar{a}_{1}e_{x}e_{y} \right \rangle.
\end{equation}
Focusing only on the sub-space spanned by two electrons in $e$ states, we can
write an explicit form of this wavefunction:
\begin{equation}
\Phi \left(^3A_2,m_{\text{s}}=+1\right) = \frac{1}{\sqrt{2}}\left[e_{x}(1)e_{y}(2)-
e_{y}(1)e_{x}(2)\right]\cdot\alpha(1)\alpha(2),
\end{equation}
where 1 and 2 is a short-hand notation for spatial or spin coordinates of the
two electrons. The electron density associated with this wavefunction is
\begin{eqnarray}
\rho(1) &=& 2\int\sum_{\text{spin}} \Phi(1,2) \Phi^{\dag}(1,2) d2 \nonumber \\
&=& \int\left[e^2_{x}(1)e^2_{y}(2)-2e_{x}(1)e_{y}(1)e_{y}(2)e_{x}(2)+e_{y}^2(1)e_{x}^2(2)\right]d2=e^2_{x}(1)+e^2_{y}(1),
\end{eqnarray}
which is expected for a single Slater determinant. Naturally, the same
applies to other values of $m_s$. Pair densities of the form
$e_{x}(1)e_{y}(1)$ do not appear in the final expression because
single-particle states $e_x$ and $e_y$ are orthogonal.

The wavefunction of the $^1A_1$ state is a sum of two Slater determinants
\cite{Doherty_NJP_2011,Maze_NJP_2011}:
\begin{equation}
\left |^1A_1\right \rangle =\frac{1}{\sqrt{2}}
\left(\left| a_{1}\bar{a}_{1}e_{x}\bar{e}_{x} \right \rangle + \left| a_{1}\bar{a}_{1}e_{y}\bar{e}_{y} \right \rangle\right).
\end{equation}
The explicit form of this wavefunction, focusing on the two electrons in the
$e$ states, is then:
\begin{equation}
\Phi \left(^1A_1\right) = \frac{1}{\sqrt{2}}\left[e_{x}(1)e_{x}(2)+e_{y}(1)e_{y}(2)\right]
\cdot\frac{1}{\sqrt{2}}\left[\alpha(1)\beta(2)-\beta(1)\alpha(2)\right],
\end{equation}
with a symmetric real-space part and an anti-symmetric spin part. The
real-space part of the wavefunction has thus a bosonic symmetry, and this
causes obvious problems when treating such wavefunctions within
density-functional theory based approaches. However, similar to the $^3A_2$
state, the electron density associated with $e$ electrons in the $^1A_1$
state can be shown to be $e^2_{x}(1)+e^2_{y}(1)$:
\begin{eqnarray}
\rho(1) &=& 2\int\sum_{\text{spin}} \Phi(1,2) \Phi^{\dag}(1,2) d2 \nonumber \\
&=& \int\left[e^2_{x}(1)e^2_{x}(2)+2e_{x}(1)e_{y}(1)e_{x}(2)e_{y}(2)+e_{y}^2(1)e_{ye}^2(2)\right]d2=e^2_{x}(1)+e^2_{y}(1),
\end{eqnarray}
Assuming that the single-particle orbitals do not change much in the two
electronic states, this implies that electron densities and thus forces on
atoms are very similar in the two electronic states as long as the $C_{3v}$
symmetry is maintained. Furthermore, in this consideration the mixing of the
$^1A_1$ singlet with higher-lying singlets (with the nominal electron
configuration $e^4$) is ignored. Since the latter has a very high energy
\cite{Ranjbar_PRB_2011}, this assumption seems justified.

The conclusion about very similar electron densities in the singlet and the
triplet state seems to be in contrast to knowledge acquired considering
Heitler-London-type wavefunctions. These occur, e.g.\, in studying the
hydrogen molecule or coupled quantum dots. In these situations electron
densities of the triplet and the singlet wavefunctions are very different due
to the appearance of mixed densities of the form $e_{x}(1)e_{y}(1)$
\cite{levine}. This is due to the fact that single-electron orbitals in the
Heitler-London wavefunction are typically not orthogonal.

A similar reasoning applies to the electron density of the $^1E$ state. In
this case the $C_{3v}$ symmetry can be broken, but the energy related to
symmetry breaking is small (see above). However, $^1E$ level can interact
with $^1E'$ level (electron configuration $a_1e^3$), and thus the resulting
electron density is slightly different from that of the $^1A_1$ state. There
are some indications that the admixing coefficient of the $a_1e^3$
configuration is about 0.3 \cite{Manson_arXiv_2010}. In broad terms, this
indicates that the equilibrium geometry of the defect in the $^1E$ state is
(approximately) a weighted average of the geometries in $^3A_2$ and $^3E$
states. This information was used in placing the total energy curve
pertaining to the $^1E$ state in Fig.\ 3.

In the main text we mention the experimental results of Robledo \emph{et al.}
\cite{Robledo_NJP_2011} and Acosta \emph{et al.} \cite{Acosta_PRB_2010} as
indicating that a small energy barrier exists for nonradiative relaxation
from the $^1E$ state. Under the assumption that the ISC is dominated by
single-phonon emission, energies of 17\,meV \cite{Robledo_NJP_2011} and 15
meV \cite{Acosta_PRB_2010} have been inferred from the temperature dependence
of the $^1E$ state lifetime. The fact that the equilibrium geometries of the
$^1E$ and $^3A_2$ states are slightly different, as discussed above, makes
non-radiative relaxation via multiphonon emission a viable alternative.
However if such a process is feasible, the energy barrier inferred from a
Mott-Seitz fit to the $^1E$ state lifetime should not be interpreted as the
energy difference between the $^3A_2$ and $^1E$ states, but as the barrier
for thermally-activated nonradiative relaxation, in close analogy to the
situation discussed in the main text. The fact that the equilibrium geometry
of the $^1E$ state is a weighted average of the geometries in $^3A_2$ and
$^3E$ states, puts a value of $\sim\!0.25$ eV (the sum of the anti-Stokes
shift and the deduced barrier) as the upper limit for the separation of the
two states. Thus, if interpreted in terms of the ISC assisted by multi-phonon
emission, the results of Refs.\ \cite{Robledo_NJP_2011} and
\cite{Acosta_PRB_2010} are consistent with a model in which only two singlet
states are invoked.

%
